\documentclass[oneside,10pt]{swissi}

\newcommand{\papertitle}{Compliant AI Infrastructure for Regulated Finance}
\newcommand{\papersubtitle}{A tiered multi-agent framework with DLT audit trails for financial operations in DACH}

\newcommand{\papercitationauthors}{Kurz, W. and Magg, R.}

\newcommand{\paperauthors}{%
  Walter Kurz\textsuperscript{1}%
  , Reinhard Magg\textsuperscript{1}%
}

\newcommand{\paperaffiliations}{%
  \textsuperscript{1}Swissi Institute for AI,
  \href{mailto:kurz@swissi-ai.institute}{kurz@swissi-ai.institute};
  \href{mailto:magg@swissi-ai.institute}{magg@swissi-ai.institute}%
}

\newcommand{\paperdeclarationaffiliations}{%
  \begin{tabular}[t]{@{}l@{}}
    \textsuperscript{1}Swissi Institute for AI,
    \href{mailto:kurz@swissi-ai.institute}{kurz@swissi-ai.institute};
    \href{mailto:magg@swissi-ai.institute}{magg@swissi-ai.institute}%
  \end{tabular}%
}

\newcommand{\papercorresponding}{%
  Walter Kurz,
  \href{mailto:kurz@swissi-ai.institute}{kurz@swissi-ai.institute},
  \href{https://orcid.org/0009-0006-8045-4775}{ORCID 0009-0006-8045-4775}%
}

\newcommand{\paperfunding}{%
  This research received no external funding.%
}
\newcommand{\paperconflictsofinterest}{%
  The authors declare no conflicts of interest.%
}
\newcommand{\paperdataavailability}{%
  The data and code supporting this article are available from the corresponding author on reasonable request.%
}
\newcommand{\paperaitools}{%
  Generative AI tools were used for editorial work only, such as language editing and formatting, in accordance with international scientific standards. The authors verified the content and remain fully responsible for the article.%
}
\newcommand{\paperauthorcontributions}{%
  All authors contributed equally to this article.%
}

\title{\papertitle \\[0.35em]
\normalsize \papersubtitle}

\author{{\fontsize{8}{10}\selectfont\paperauthors}\\[0.9em]
{\scriptsize\paperaffiliations}}
\date{\vspace{1.5em}}

\begin{document}
\maketitle

\begin{abstract}
We present a compliance-first architecture for AI in regulated finance that treats regulation as an orientation layer rather than a deterministic ruleset. A matrix of regulatory intent and exposure provides a compact classification handle, which a governed policy compiler then maps into concrete prohibitions, obligations and runtime budgets. Prohibitions constrain feasibility and block externalisation, while obligations extend tasks with artefacts that must meet explicit admissibility criteria. Committee activation remains policy-driven and proportionate, preserving efficiency while ensuring supervisory oversight. Evidence, decisions and reason codes are bound to a permissioned DAG with deterministic timestamping, enabling replay, provenance checks and clear attribution of failure. Clause-level legal indexing with effective dates and capability-based agent routing ensure portability across DACH and the wider EU. The result is assurance by construction: compliance is embedded in execution and verifiable by auditors without sacrificing proportionality or transparency.
\end{abstract}

\keywords{DACH finance; regulated financial institutions; multi agent expert system; policy compiled orchestration; objective under constraints; permissioned DLT; DAG timestamping; audit trails; EU AI Act; MiFID II; DORA; GDPR; human oversight; execution gating; ESG budgets; verification and assurance}

\bigskip

\section{Introduction}
The financial industry combines data intensity, complex decision processes and strict regulatory oversight, making it a natural environment for the application of artificial intelligence. Yet the adoption of AI in this domain is constrained by the absence of architectures that meet supervisory expectations for accountability, resilience and auditability. Existing implementations often prioritise performance or innovation at the expense of compliance, leaving institutions without frameworks that can be both operationally effective and regulatorily sound. This study addresses that gap by proposing a reference architecture for compliant AI in finance, designed to integrate supervisory requirements into system design from the outset. The analysis focuses on financial institutions in the DACH region. Germany, Austria and Switzerland combine deep capital markets, dense supervisory practice and high cross-border integration \cite{ecbFISEA2024,bisIBS2025,sifEU2025}. Institutions operate under comparable prudential expectations and reporting cultures: Germany and Austria within the EU prudential framework, Switzerland through equivalence-based alignment in selected domains. This convergence provides a coherent testing ground, supporting methodological consistency while enabling meaningful comparison across jurisdictions \cite{crr2013,crd2013,its2021451,ec_equivalence,sifEU2025}. For the purposes of this study, financial institutions are defined as supervised entities that accept deposits or other repayable funds, extend credit or investment services, insure or reinsure risk, manage collective assets, operate payment or settlement systems, or run market infrastructure. \cite{crr2013,mifid2_2014,solvency2_2009,ucits_2009,aifmd_2011,psd2_2015,emir_2012,csdr_2014,mifir_2014}. Within the prudential perimeter, the taxonomy distinguishes credit institutions and banks under CRR/CRD, investment firms under MiFID~II, insurance and reinsurance under Solvency~II, asset managers under UCITS and AIFMD, and market infrastructures such as trading venues, CCPs and CSDs under MiFIR, EMIR and CSDR. Payment and e-money institutions fall under PSD and EMD \cite{crr2013,crd2013,mifid2_2014,mifir_2014,solvency2_2009,emir_2012,csdr_2014,psd2_2015,emd2_2009,ucits_2009,aifmd_2011}. A functional taxonomy complements this perimeter by grouping front-office decisioning and advisory, treasury and risk, payments and settlement, compliance and reporting, and back-office operations. Outsourcing and critical ICT providers enter scope where the ICT risk framework applies \cite{dora_2022}. European financial services are governed by an extensive regulatory corpus. The EU \emph{acquis} sets binding requirements on prudential soundness, market conduct and investor protection through CRR/CRD, MiFID~II and MiFIR, Solvency~II, EMIR and CSDR, PSD and EMD, AML directives, MiCA, DORA and GDPR \cite{crr2013,crd2013,mifid2_2014,mifir_2014,solvency2_2009,emir_2012,csdr_2014,psd2_2015,emd2_2009,amld4_2015,amld5_2018,mica_2023,dora_2022,gdpr_2016}. These instruments specify licensing, governance, outsourcing, logging, testing, resilience and data protection duties, and enable passporting across the single market \cite{crd2013,esmaArt16,delReg2017_565,dora_2022,gdpr_2016,esmaArt34,ebaPSD2Art28}. Supervised firms must evidence compliance through internal control systems, audit trails and supervisory reporting \cite{solvency2_2009,esmaArt16,delReg2017_565,its2021451}. The DACH region presents a coherent yet diverse regulatory landscape. Germany applies EU law through national instruments such as the \emph{Kreditwesengesetz} (KWG) and \emph{Wertpapierhandelsgesetz} (WpHG) under BaFin supervision \cite{germany_kwg,germany_wphg,bafin_wphg}. Austria implements EU law through the \emph{Bankwesengesetz} (BWG) and \emph{Wertpapieraufsichtsgesetz} 2018 (WAG~2018) under FMA supervision \cite{austria_bwg,austria_wag2018,fmaIntegrated}. Switzerland enforces FinSA and FinIA with FINMA, framed by EU equivalence decisions and bilateral arrangements \cite{fedlex_finsa,fedlex_finia,finmaLegalBasis,ec_equivalence,sifEU2025}. The region’s institutional scale and diversity are material: universal banks, savings and cooperative networks, global insurers, specialised asset managers and market utilities operate under comparable supervisory expectations, while preserving sufficient heterogeneity to test generality \cite{ecbFISEA2024,efamaFactbook2025}. Burden asymmetries persist: in Austria, smaller firms bear proportionally higher fixed compliance costs, with studies documenting that regulatory obligations weigh more heavily on them than on larger institutions \cite{eibis2024Austria,fsb2019SME,bisfsi2018Proportionality,Kurz2025FMAAustria}.

\section{Related Work}

Prior literature on regulated finance in Europe has largely examined prudential soundness, market conduct, and investor protection under the comprehensive EU \emph{acquis}, with attention to national transpositions in Germany and Austria and equivalence-based coordination with Switzerland. Governance frameworks for financial technology have increasingly incorporated operational risk and systemic resilience, yet compliance in practice is often confined to procedural assurances rather than embedded as an architectural invariant. Research on multi-agent systems in finance has advanced considerably, particularly in algorithmic trading, portfolio optimisation, and credit risk modelling. Examples include hierarchical deep reinforcement learning for portfolio optimisation \cite{Sun2025MultiAgentHDRL}, multi-agent dynamic portfolio learning systems \cite{Cheng2024MultiagentRL}, and surveys of reinforcement learning applications in finance \cite{Hambly2021RLFinance,Malibari2023RLFintech}.  These approaches typically optimise task-level performance under static policy assumptions and rarely integrate verifiable execution evidence as part of the architecture itself. Parallel work on enterprise logging and model risk management underscores the need for traceability, oversight, and effective challenge. Surveys of audit log security highlight difficulties in ensuring integrity and immutability \cite{Ali2021AuditLogSurvey}, while research on BPM systems shows how timestamped, tamper-evident logs can support transparent audit regimes \cite{IJITMIS2025BPMAudit}. Yet cryptographic timestamping is seldom treated as a first-class architectural principle, and its integration with multi-agent orchestration remains limited. Distributed ledger research in financial services has concentrated on post-trade settlement, asset tokenisation, and shared record-keeping. Permissioned blockchain ecosystems can improve transparency and enable authorised audit access \cite{Faccia2022PermissionedAudit}, and recent studies link blockchain with AI to strengthen audit quality \cite{Arham2025AIBlockchainAudit}. Still, audit-grade logging via directed acyclic graph (DAG) architectures for AI coordination remains under-explored. Empirical analyses of regulatory burden further demonstrate the case for automation and proportionality. Studies of the Austrian perimeter show that compliance overhead scales disproportionately for SMEs, motivating system designs that embed automated evidence generation and allow for proportional controls \cite{Kurz2025FMAAustria}. Within ICT risk regulation, the Digital Operational Resilience Act (DORA) stresses robust risk management, third-party oversight, resilience testing, and logging requirements \cite{DORA_EIOPA2025,DORA_Skadden2024,EBA_RTS2024}. These reinforce the need for architectures that decouple verifiable orchestration from model training, enforce strict role segregation, and support dual-control governance. Taken together, these strands highlight a persistent gap: no existing enterprise architecture simultaneously treats compliance as both a compile-time and run-time constraint, integrates multi-agent coordination under gated execution, and generates supervision-ready evidence through deterministic timestamping and inclusion proofs. This gap is particularly salient for tiered designs that must serve both institutional and consumer interfaces while remaining aligned with DACH-specific regulatory norms.

\section{Contribution}

Current compliance mechanisms in financial institutions remain predominantly ex-post: obligations are verified through periodic audits, reconciliations, and manual attestations, while system execution itself proceeds without embedded regulatory constraints. This approach creates structural gaps, since audit trails can be incomplete, supervisory checks are delayed, and interpretations of legal texts remain siloed across departments. Controls are reactive rather than preventative, and compliance risk accumulates in the time between transaction and inspection. Addressing these deficiencies requires a design in which regulatory duties are formalised as machine-checkable constraints and enforced as part of orchestration and execution rather than appended as a reporting layer. This research develops an enterprise reference architecture for a tiered, compliant AI system in finance, aligned with the regulatory taxonomy and supervisory context of the DACH region yet transferable across the EU. The architecture encodes regulatory duties as machine-checkable constraints and binds them to orchestration and execution. A policy store and rule engine compile EU and DACH requirements into checks enforced both before and during execution. Multi-agent coordination is role scoped across institutional and consumer tiers, with explicit permissions, explainability thresholds and segregation of duties. Gated execution ensures that sensitive actions, such as order placement, occur only under dual control, and every step produces audit-ready evidence. Verification is achieved through a permissioned distributed ledger that employs a directed acyclic graph (DAG) for deterministic timestamping and inclusion proofs. The ledger captures routing instructions, budget usage and control attestations without imposing settlement semantics on application workflows. Decisions are formulated as objectives under constraints, with runtime budgets for risk, latency and ESG, enabling transparent trade-offs at schedule time. The architecture remains technology-agnostic at model level and interoperable with existing systems via supervised queues and APIs. In this way, it provides a reference design that meets supervisory expectations in DACH while supporting broader applicability across EU jurisdictions.

\section{System Architecture}

The proposed system architecture embeds regulatory compliance as a first-class design primitive rather than an external audit layer added post hoc. Orchestration and execution are bounded by machine-enforceable constraints compiled directly from legal provisions, ensuring that system behaviour remains verifiable at run time. The architecture is tiered across institutional and consumer contexts, portable across jurisdictions, and agnostic to specific model technologies. Its emphasis lies on properties that supervisory authorities recognise as audit-ready: determinism of execution, separation of roles, and transparent evidence generation. A distinctive element of the design is the integration of a permissioned distributed ledger that employs a directed acyclic graph (DAG) structure to anchor audit trails. Each agent action, constraint evaluation, and gating decision is recorded as a tamper-evident event with deterministic timestamping and inclusion proofs. Unlike traditional logging, which remains vulnerable to post-processing or selective disclosure, the DAG ledger produces an immutable sequence of verifiable attestations that can be inspected by internal control functions and supervisory authorities. The ledger is deliberately lightweight: it does not impose settlement semantics on application workflows but serves solely as a compliance substrate, capturing routing instructions, evidence packs, and oversight signatures. This enables continuous assurance without altering financial transaction flows.

\subsection{Agnostic Regulatory Intent and Exposure Matrix}

\begin{figure}[H]
  \centering
  \includegraphics[width=\linewidth]{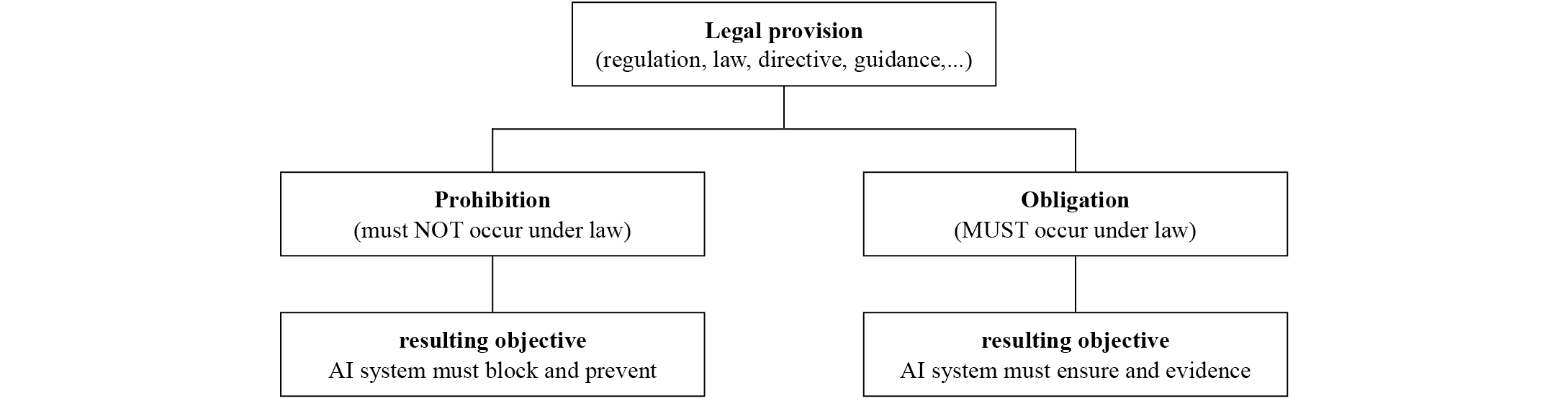}
  \caption{Taxonomy of regulatory intent. Each legal provision (regulation, law, directive, guidance) maps to Prohibition or Obligation, indicating the primary reasoning objective.}
  \label{fig:rim-taxonomy}
\end{figure}

European financial regulation is dense and layered across prudential, conduct, market integrity, data protection and ICT risk instruments. Designing an enterprise system by listing acts and articles provides little direct operational guidance. To address this, we introduce a regulation-agnostic taxonomy that functions as a design primitive. Its purpose is to translate legal text into machine-enforceable control functions that govern orchestration and execution while remaining portable across DACH and the wider EU. We adopt a functional reduction of legal provisions to their operational effect on system behaviour: either a rule forbids an outcome or it demands an outcome with evidence. Compliance therefore reduces to abstention from a prohibited act or to the performance of a positive act that leaves a verifiable trace. This yields a law- and regulation-agnostic orientation in which agents need only decide whether to avoid or to act with evidence, while the specific statutory text is compiled at a later stage into guard families and evidence templates, with exposure handled separately for internal versus external outputs. Practically, this agnostic treatment simplifies handling heterogeneous regulations because the agent can synthesise, per task and context, a checklist of prohibitions and obligations that orchestration executes and tracks, with each item either preventing externalisation or requiring a verifiable artefact with provenance, deterministic timestamping and DAG inclusion proof; alternatively a dedicated compliance agent can apply the same checklist as a pre-output gate, sign the evidence pack, and authorise or withhold release.\footnote{A prohibition can be illustrated with a consumer-tier agent that is not permitted to instruct a client to purchase a particular security or to transmit an order on their behalf without their permission. In this case the guard blocks both recommendation and order externalisation, recording at most a hold or release attestation on the DAG ledger. An obligation is exemplified by onboarding procedures, which require identification and know-your-customer documentation. Here the workflow executes the prescribed checks and generates a signed artefact with provenance metadata, deterministic timestamping, and a DAG inclusion proof before any advisory or execution services can be activated.} The taxonomy is formalised as a two-dimensional schema, the \emph{Regulatory Intent and Exposure Matrix}. Each legal provision receives two labels \emph{rule type}, capturing intent as either Prohibition or Obligation, and \emph{exposure}, capturing context as Internal or External.
\begin{equation}
v = (t,e), \quad t \in \{\text{Prohibition}, \text{Obligation}\}, \; e \in \{\text{Internal}, \text{External}\}
\label{eq:rim-label}
\end{equation}
The matrix is an orientation device rather than an execution rule. The label \(v=(t,e)\) expresses a stance that guides reasoning and proposes candidate guard families and evidence templates; the selection and activation of concrete gates is compiled later by the rule engine from the policy store and context features, and may be revised by human oversight. In short, \(v\) narrows the search space; it does not by itself prescribe what the system must do. A compact label provides the canonical representation. This schema enables candidate guard families and evidence templates to be proposed without first naming a statute, ensuring portability across instruments and jurisdictions. Prohibitions mark contexts where the system should consider blocking externalisation; obligations mark contexts where the system should consider requiring a positive act and admissible evidence. Whether these candidates become active gates depends on compiled policy, context, and approvals, not on the matrix alone.
This binary mapping is used for orientation in reasoning and as a compact handle for later compilation, as shown in Figure~\ref{fig:rim-taxonomy}.

Figure~\ref{fig:rim-taxonomy} frames two reasoning stances rather than execution rules. A Prohibition prompts scrutiny of potentially disallowed behaviour and a hold on externalisation until resolved. An Obligation prompts identification of duties and anticipation of evidence likely required if the task proceeds. Clauses are stored as atomic statements with mapping and citation to ensure fast reading and consistent interpretation. Concrete gates and logging are derived later once reasoning has reached sufficient confidence.

\begin{figure}[H]
  \centering
  \includegraphics[width=\textwidth]{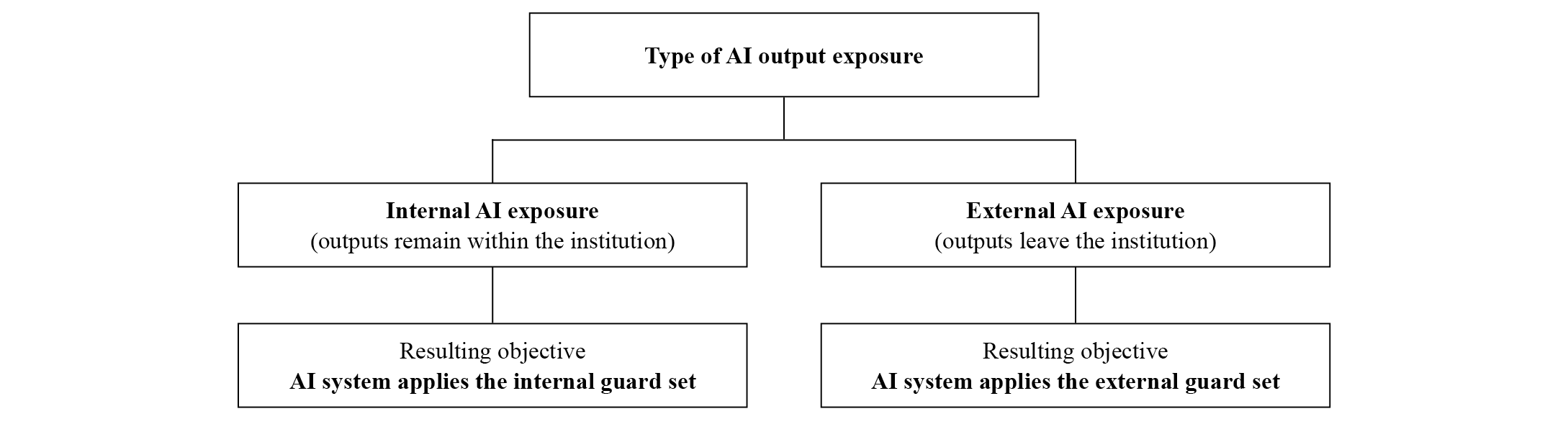}
  \caption{Taxonomy of exposure. Internal indicates outputs confined to the institution; external indicates outputs leaving the institutional boundary. Mixed workflows default to external.}
  \label{fig:exposure-taxonomy}
\end{figure}

The second axis classifies exposure. Internal exposure refers to outputs confined to the institution, while external exposure refers to outputs delivered to clients, counterparties, authorities or markets. The distinction is contextual and combines with intent in Figure~\ref{fig:exposure-taxonomy}.

\begin{figure}[H]
  \centering
  \includegraphics[width=\linewidth]{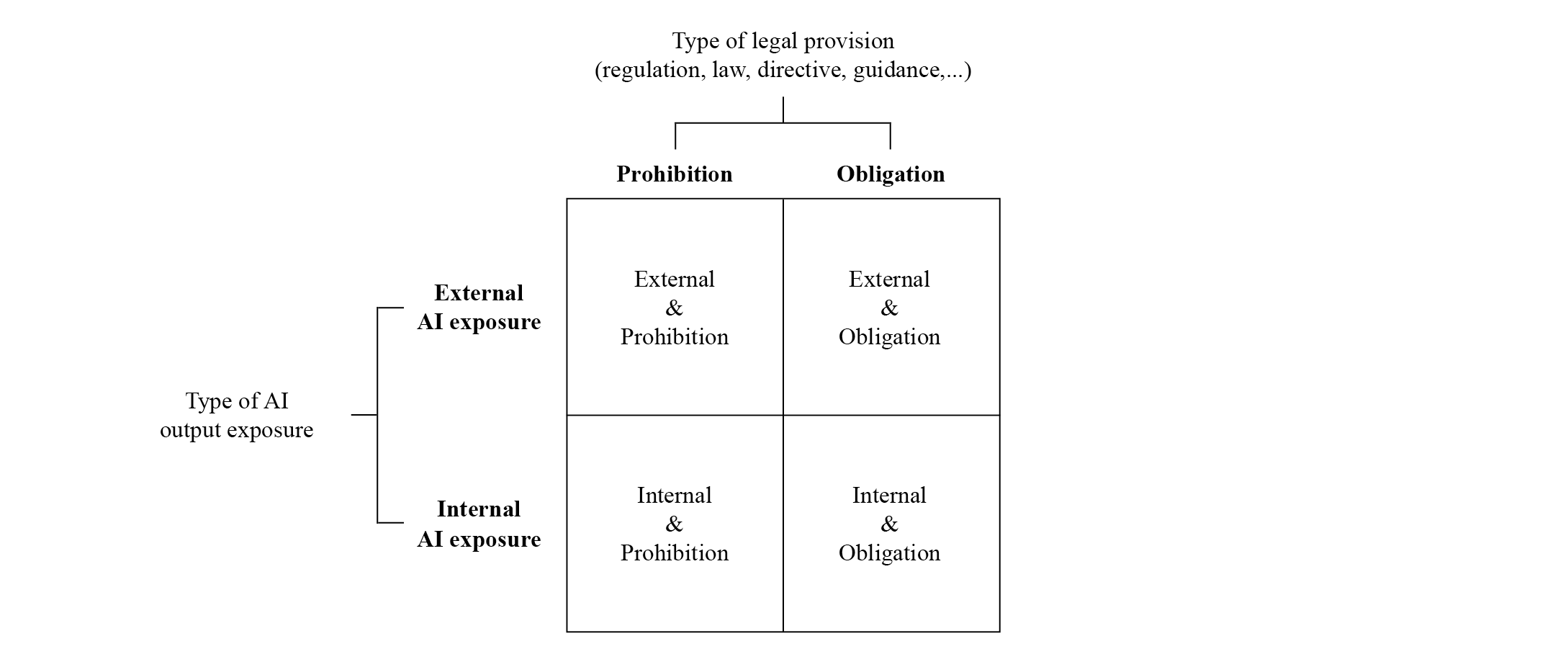}
  \caption{Regulatory intent and exposure matrix. Quadrants orient reasoning and suggest candidate guard families and evidence expectations. Mixed workflows default to external.}
  \label{fig:rim-matrix}
\end{figure}

This classification orients reasoning by establishing the destination of outputs. Once combined with intent, it guides which questions and candidate actions are relevant. Figure~\ref{fig:rim-matrix} presents the joint view.

At run time an agent performs a lightweight context check against the proposed matrix. If no mapped provision appears to apply, the agent continues under general IT controls and ordinary policy. If the check suggests applicability, the agent classifies the context using the matrix labels and treats that label as an orientation handle for reasoning rather than as an enforcement command, subject to revision as scope and evidence evolve. The context check can be formalised as a binary predicate that determines whether a task falls under a mapped legal provision:
\begin{equation}
\chi:\mathcal{T}\to\{0,1\},\qquad
\chi(\tau)=
\begin{cases}
1 & \text{if the mapped provision applies to }\tau\\
0 & \text{otherwise}
\end{cases}
\label{eq:context-check}
\end{equation}
If \(\chi(\tau) = 0\), the task proceeds under baseline IT controls and operational policy. If \(\chi(\tau) = 1\), the task is assigned a regulatory label \(v=(t,e)\) as defined above. This label serves as an orientation handle that anchors subsequent reasoning.

Once classified, the agent frames the task as a constrained optimisation problem of the form
\begin{equation}
\begin{aligned}
\max_{\pi \in \Pi} \quad & U(\pi,\tau) \\
\text{s.t.} \quad & C(\pi,\tau,v) = \text{true},
\end{aligned}
\label{eq:optimisation}
\end{equation}
where \(\pi\) is a candidate policy, \(U(\pi,\tau)\) denotes the utility of executing \(\pi\) on task \(\tau\), and \(C(\cdot)\) encodes the selected guard families and evidence checks compiled from \(v\) and context; the matrix label \(v\) orients this compilation but does not, by itself, determine execution. Prohibitions are represented as hard constraints considered for activation to prevent disallowed outcomes, while obligations are represented as candidate duties that, if selected by policy, require admissible evidence. If no provision applies (\(\chi(\tau)=0\)), the constraint set \(C\) collapses to baseline IT controls and ordinary operational policy. This formulation connects the taxonomy to decision-making without collapsing orientation into determination.

Minting the classification, checklist and gating decisions to the DAG enables ex post assurance and explicit failure analysis: an auditor can reconstruct the task state \(\phi(\tau)\) (see \eqref{eq:qr-candidates}), recompute \(\chi(\tau)\) (see \eqref{eq:context-check}) and the label \(v=(t,e)\) (see \eqref{eq:rim-label}) to test whether the agent misunderstood the context, verify that each selected guard \(C(\pi,\tau,v)\) in the optimisation constraint (see \eqref{eq:optimisation}) was satisfied at the moment of externalisation, and match every obligation item to a concrete artefact with provenance, deterministic timestamping and inclusion proof; this makes distinct failure modes observable, including (i) misclassification of intent or exposure, (ii) correct classification and checklist generation but incomplete execution, (iii) execution with insufficient evidence quality or missing signatures, and (iv) sequencing or gating errors that allowed externalisation without preconditions, as well as budget breaches for risk, latency or ESG. The proposal turns compliance from after-the-fact attestation into assurance-by-construction, where reasoning, decisions and evidence are bound cryptographically so reviewers can check not only outcomes but also whether the agent’s understanding and follow-through were correct.

We propose a short, curated aid for cases where the context check indicates applicability. A quick reference vector is a concise entry prepared by a qualified oversight committee; it orients reasoning without prescribing execution. Each entry names the action archetype, states the typical matrix label, notes salient triggers and the initial evidence posture, and gives an escalation contact. Entries are versioned and signed so that provenance remains clear; they can be revised as scope or information changes. When a task matches such an entry, the agent adopts the entry as orientation and continues analysing the task toward the decision objective; if no entry fits, the agent proceeds under baseline controls and may flag the gap for later inclusion. The intent is speed and consistency in judgement, not hard rules.
\begin{align}\label{eq:qr-vector}
r = (a, v, \kappa, \epsilon, \nu, \sigma)
\end{align}
where \(a\) is the action archetype, \(v=(t,e)\) is the matrix label from Equation~\eqref{eq:rim-label}, \(\kappa\) captures salient triggers and context features, \(\epsilon\) records the initial evidence posture, \(\nu\) denotes version and effective interval, and \(\sigma\) is the oversight signature.
\begin{align}\label{eq:qr-candidates}
R(\tau) = \{\, r \in Q \mid \mathrm{sim}\big(\kappa_r,\;\phi(\tau)\big) \ge \theta \,\}
\end{align}
where \(Q\) is the approved set, \(\phi(\tau)\) is a compact feature map of the task, and \(\theta\) is a reviewable threshold. If \(R(\tau)=\varnothing\), the task proceeds under baseline controls and the context check is recorded for later improvement. If \(R(\tau)\neq\varnothing\), the top-ranked entry \(\hat r\) is adopted as orientation and the analysis continues toward the decision objective in Equation~\eqref{eq:optimisation}.

\subsection{Prohibitions as feasibility and obligations as task extension}

The matrix provides orientation only; concrete enforcement is selected later by policy compilation and may be revised by human oversight. Given the label \(v=(t,e)\) from \eqref{eq:rim-label} and the context determined by the predicate \(\chi(\tau)\) in \eqref{eq:context-check}, the rule engine surfaces candidate prohibitions \(P\) and obligations \(O\), then selects active subsets \(P^{\star}\subseteq P\) and \(O^{\star}\subseteq O\) together with a strict partial order \(\prec\) over \(P^{\star}\cup O^{\star}\). The order \(\prec\) permits short-circuit evaluation where appropriate and defines a narrow top tier \(P^{\perp}\subseteq P^{\star}\) of dominant disqualifiers. Obligations map to mandatory sub-actions and a verifiable evidence artefact; let \(E(\cdot)\) denote the admissibility predicate capturing provenance, signatures, deterministic timestamping and DAG inclusion proof. None of these selections is implied by the matrix itself; the matrix narrows the search space, the policy store determines activation.

Prohibitions are encoded as feasibility constraints that prevent disallowed outcomes, while obligations extend the task with preconditions that must hold before optimisation under \eqref{eq:optimisation} is meaningful. The feasible policy set for task \(\tau\) is
\begin{equation}
\Pi_{\mathrm{feas}}(\tau)=\Bigl\{\pi\in\Pi \ \Big|\ \bigwedge_{p\in P^{\star}} \neg p(\pi,\tau)\ \wedge\ \bigwedge_{o\in O^{\star}} E\bigl(\mathrm{artefact}_o(\tau)\bigr)\Bigr\}.
\label{eq:feasible}
\end{equation}
Optimisation proceeds lexicographically to respect feasibility first and utility second,
\begin{equation}
\pi^\star \in \arg\max_{\pi\in\Pi_{\mathrm{feas}}(\tau)} U(\pi,\tau)
\label{eq:lexi}
\end{equation}

with optional secondary terms in \(U\) for obligation quality such as latency, cost or completeness. This construction preserves the law- and regulation-agnostic stance: prohibitions shape the admissible set, obligations extend the task graph and demand admissible artefacts, and any burden from satisfying obligations may then be traded within the objective once feasibility is secured.

Externalisation refers to any output that leaves the institutional boundary, including recommendations, orders, client-facing summaries, supervisory reports and API calls, whereas internal artefacts, logs and operator messages remain non-external unless later reused. The gate outcome formalises block versus release for a proposed output \(y(\pi,\tau)\),
\begin{equation}
\mathrm{Externalise}(\pi,\tau)=
\begin{cases}
\textsc{deny}(c) & \text{if }\exists p\in P^{\star}\text{ with }p(\pi,\tau)=1\ \ \text{or}\ \ \exists o\in O^{\star}\text{ with }E(\mathrm{artefact}_o(\tau))=\text{false},\\[4pt]
\textsc{allow}(y) & \text{otherwise,}
\end{cases}
\label{eq:externalise}
\end{equation}
where \(c\) is a machine- and human-readable reason code minted to the DAG with a hold or release attestation. Blocking does not suppress internal reasoning; the system returns an explicit explanation to operators and records the decision and grounds for audit.

Short-circuit evaluation reduces latency and user friction without sacrificing assurance. If there exists a dominant disqualifier \(p\in P^{\perp}\) with \(p(\pi,\tau)=1\), then \(\mathrm{Externalise}(\pi,\tau)=\textsc{deny}(c)\) by \eqref{eq:externalise} and the remaining lower-tier checks in \(P^{\star}\cup O^{\star}\) are marked \emph{skipped by policy}. The audit pack minted to the DAG includes the orientation set, the selected \(P^{\star},O^{\star}\), the order \(\prec\), the firing guard in \(P^{\perp}\), the skipped set with its policy basis, all artefact hashes and admissibility outcomes \(E(\cdot)\), and the reason code \(c\). This enables an auditor to reconstruct the state \(\phi(\tau)\) used in the quick-reference matching \eqref{eq:qr-candidates}, to verify that the selected guards \(C(\pi,\tau,v)\) in the optimisation constraint \eqref{eq:optimisation} were satisfied at the moment of externalisation, and to distinguish failure modes such as misclassification, incomplete execution of obligations, insufficient evidence quality, or sequencing errors that allowed externalisation without preconditions.

A composite example illustrates the semantics. Consider a cross-border retail transaction where the amount exceeds the threshold for retail investors and the client is under age. Orientation via \(v\) surfaces multiple candidates; policy selects \(P^{\star}\) that includes an under-age prohibition and places it in \(P^{\perp}\) under \(\prec\). The under-age guard fires, so \(\mathrm{Externalise}(\pi,\tau)=\textsc{deny}(c)\) by \eqref{eq:externalise}; cross-border and amount checks are skipped by policy with the dominance certificate recorded in the audit pack. If, in a different context, no dominant disqualifier fires, obligations such as identification and know-your-customer documentation become active preconditions, and externalisation is permitted only once admissibility \(E(\cdot)\) holds, after which utility is optimised over \(\Pi_{\mathrm{feas}}(\tau)\) according to \eqref{eq:lexi}. In both cases the matrix remains an orientation device; selection and activation are determined by compiled policy and oversight, and the full reasoning trail is minted to the DAG for ex post assurance.

We assume only mild regularity, namely that dominant disqualifiers in \(P^{\perp}\) remain policy invariant for a given task state, so that \(p(\pi,\tau)=p(\pi',\tau)\) holds for all \(\pi,\pi'\in\Pi\); in practice these include conditions such as age thresholds, sanctions hits, or bans tied to specific product tiers. The admissibility predicate \(E(\cdot)\) is used to capture evidence quality requirements, including provenance, signatures, deterministic timestamping, and DAG inclusion, and is evaluated directly on the artefacts compiled for \(\tau\).

Under these assumptions the feasible set is monotone: if the selected sets expand such that \(P^{\star}_1\subseteq P^{\star}_2\) and \(O^{\star}_1\subseteq O^{\star}_2\), then it follows that \(\Pi_{\mathrm{feas}}^{(2)}(\tau)\subseteq \Pi_{\mathrm{feas}}^{(1)}(\tau)\), where \(\Pi_{\mathrm{feas}}^{(k)}\) is defined by \eqref{eq:feasible} under the selection \(P^{\star}_k,O^{\star}_k\). The result follows immediately from \eqref{eq:feasible}, since adding further prohibitions or obligations introduces additional conjuncts that can only reduce the satisfying set.

Short-circuit evaluation is sound whenever dominant checks are policy invariant: if a single \(p\in P^{\perp}\) evaluates to one for some \(\pi\), then it evaluates to one for all \(\pi'\), which implies \(\Pi_{\mathrm{feas}}(\tau)=\varnothing\) by \eqref{eq:feasible} and forces \(\mathrm{Externalise}(\pi,\tau)=\textsc{deny}(c)\) by \eqref{eq:externalise}. Conversely, if no element of \(P^{\perp}\) fires and all selected obligations satisfy admissibility so that \(E(\cdot)=\text{true}\), then \(\Pi_{\mathrm{feas}}(\tau)\) remains nonempty and the optimisation problem in \eqref{eq:lexi} is well defined. Evaluation both terminates and remains auditable. Since the sets \(P^{\star}\cup O^{\star}\) are finite and \(\prec\) is a strict partial order, a topological evaluation sequence necessarily exists, with short-circuiting at \(P^{\perp}\) further bounding runtime. The DAG record contains the orientation set, the selected \(P^{\star},O^{\star}\), the order \(\prec\), any firing element of \(P^{\perp}\), the admissibility outcomes \(E(\cdot)\), and the resulting gate decision \eqref{eq:externalise}, which together enable replay and independent re-evaluation using \eqref{eq:context-check}, \eqref{eq:rim-label}, \eqref{eq:feasible} and \eqref{eq:optimisation}.

If obligation quality is intended to influence the objective once feasibility has been established, the framework admits a straightforward refinement in which \(U(\pi,\tau)=U_0(\pi,\tau)-\lambda^\top q(\tau)\), with \(q(\tau)\) aggregating latency, cost, and completeness across satisfied obligations and \(\lambda\ge 0\) denoting policy weights. The lexicographic regime in \eqref{eq:lexi} guarantees that such trade-offs are only evaluated within the feasible set \(\Pi_{\mathrm{feas}}(\tau)\).

\subsection{Agent activation by intent and exposure}

The orientation matrix provides the initial signal, but the selection and activation of executors is determined by compiled policy and remains subject to oversight revision. Institutions may maintain standing rosters of domain-specific agents and persistent committees for recurring functions, while reserving the possibility of ad hoc committees when context requires. In a banking environment, a payments roster typically spans functions such as wiring funds between accounts, executing foreign-exchange conversions, and arranging treasury transfers. For an investment firm under MiFID II, routine activities encompass the reception and transmission of client orders, order execution on behalf of clients, the provision of investment advice with suitability and appropriateness checks, portfolio management, and post-trade reporting. These rosters remain dormant until activated by the orientation label \(v=(t,e)\) from \eqref{eq:rim-label} in combination with the context check \(\chi(\tau)\) defined in \eqref{eq:context-check}. Committee formation follows a policy-driven logic. Where \(\chi(\tau)=0\) or the task is assessed as internal and low risk, such as drafting an internal email or scheduling a meeting, a single role-scoped agent executes under baseline IT controls. When \(\chi(\tau)=1\), activation depends on the exposure level \(e\), the number and dominance of active checks, and the availability of runtime budgets. This is formalised as
\begin{equation}
\mathrm{Committee}(\tau)=
\begin{cases}
S(\tau,v,\text{context}) & \text{if } \chi(\tau)=1 \land C(\tau),\\[4pt]
\{\text{primary agent}\} & \text{otherwise,}
\end{cases}
\label{eq:committee}
\end{equation}
where
\begin{equation}
C(\tau):=\big(e=\text{External}\big)\ \lor\ \big(|P^{\star}|+|O^{\star}|\ge k\big)\ \lor\ \big(P^{\perp}\neq\varnothing\big)\ \lor\ \text{budget breach (risk, latency, or cost)}.
\end{equation}
Here \(S(\cdot)\) yields the minimal role set required for the task, \(P^{\star},O^{\star}\) denote the selected checks defined in the previous subsection, \(P^{\perp}\) designates the dominant disqualifiers, and \(k\) is a policy-defined threshold. The rule ensures proportionality and avoids unnecessary coordination overhead.

Execution proceeds through a structured handshake. The orchestration agent forwards the task \(\tau\) together with the orientation \(v\) to the designated agent set or committee, which undertakes its domain-specific responsibilities and returns artefacts, rationales, and provisional outcomes. A dedicated compliance agent then applies the pre-output gate defined in \eqref{eq:externalise}; if admissibility is confirmed for all selected obligations and no prohibition fires, the output is released, whereas any failure results in blocked externalisation accompanied by a reason code. Each stage mints a committee token to the DAG ledger, recording membership, role scopes, selected checks, reason codes, and inclusion proofs, which collectively support later reconstruction and independent re-evaluation under \eqref{eq:optimisation}.

Examples illustrate proportionality and dominance. A domestic, low-value internal transfer with \(\chi(\tau)=0\) is executed by a single payments agent without committee escalation, reflecting the proportionality condition that internal exposure with few selected checks below the policy threshold \(k\) activates only the primary role. A cross-border retail securities order with \(\chi(\tau)=1\) and \(e=\text{External}\) activates the standing “securities orders” set comprising client categorisation and suitability, order execution, best-execution monitoring, and transaction reporting; if a dominant disqualifier in \(P^{\perp}\), such as an under-age client, fires, the committee collapses to the compliance gate and the externalisation returns \(\textsc{deny}(c)\) under \eqref{eq:externalise}, with all skipped checks logged as \emph{skipped by policy}. Formally,
\begin{equation}
P^{\perp}\neq\varnothing \ \wedge\ \exists p\in P^{\perp}: p(\pi,\tau)=1
\ \Longrightarrow\
\mathrm{Committee}(\tau)=\{\text{compliance gate}\},
\label{eq:dominance}
\end{equation}
which ensures that functional agents are excluded by policy while a reason-coded denial is minted under \eqref{eq:externalise}. In both directions, whether escalation occurs because exposure or policy thresholds demand it, or collapse follows from dominance being triggered, the orientation matrix identifies relevant agents while compiled policy and supervisory oversight determine actual activation.

\subsection{Legal corpus indexing and agent routing}

The legal corpus is treated as a structured, versioned asset rather than a flat vector space. Each instrument is normalised, segmented, and cited at clause level, with every chunk \(l\) carrying canonical metadata \(\mathrm{cite}(l)=\langle\text{jurisdiction},\text{instrument\_id},\text{article},\text{paragraph},\nu\rangle\), where \(\nu\) encodes version and effective interval, supplemented by tags for instrument class (prudential, conduct, data protection, ICT), intent and exposure labels, and explicit cross-references. Ingestion preserves both the original language and a controlled translation, while recording equivalence mappings between recitals, articles, RTS/ITS, and supervisory Q\&A, so that downstream retrieval can resolve different references to a single, time-scoped citation.

Indexing is hybrid in order to preserve meaning as well as citations. Three coordinated views are maintained for the same chunks: a dense semantic index for embeddings \(\phi(l)\), a sparse citation and keyword index for exact retrieval, and a citation graph \(G\) linking instruments through “refers to,” “implements,” and “interprets” edges. Queries are constructed from task features \(\psi(\tau)\), the orientation \(v=(t,e)\) from \eqref{eq:rim-label}, triggers \(\kappa\) from the quick-reference vector \eqref{eq:qr-vector}, and the exposure \(e\). Retrieval proceeds in two stages: a semantic pre-filter selects top-\(k\) candidates \(\tilde L(\tau)\) by similarity \(\mathrm{sim}(\psi(\tau),\phi(l))\) under jurisdiction and effective-date filters, and a reranker then fuses sparse scores with graph proximity to produce the final set \(L(\tau)\). Every returned item includes \(\mathrm{cite}(l)\) together with its text span, ensuring human-readable grounding.

Clustering is dual to reflect legal reading practice. We compute instrument-anchored clusters \(C_{\text{inst}}\) for each statute or linked set of acts, and aspect clusters \(C_{\text{asp}}\) derived from the citation graph and tag distributions, covering recurring themes such as suitability, client categorisation, personal recommendations, onboarding, and record-keeping. Chunks may appear in both views. Aspect clusters allow rapid coverage across instruments, while instrument clusters ensure that the exact article and paragraph remain available for citation and audit. The design avoids reliance on a single global cluster.

Routing is performed by capability rather than by statute. Institutions register domain agents with declared capabilities and evidence schemas, covering for example payments, FX, treasury, onboarding, suitability and appropriateness, personal recommendations, order execution, reporting, data protection, and ICT controls. Orientation \(v\), exposure \(e\), and the retrieved set \(L(\tau)\) jointly determine which capability bundles are relevant. The router applies the same selection logic as \eqref{eq:committee} to activate either a single role or a minimal committee. A dedicated citation clerk agent attaches \(\mathrm{cite}(\cdot)\) and extracts the governing spans, while a compliance agent executes the pre-output gate \eqref{eq:externalise}. This design keeps knowledge current and avoids brittle “one agent per law” silos.

Correlation between chunks and their governing law is guaranteed through metadata rather than embeddings. Each chunk carries its canonical citation, effective dates, and instrument identifiers (e.g. CELEX for EU acts, BaFin circular IDs, FINMA RS numbers). When retrieval proposes candidates, a citation-consistency check eliminates items whose \(\mathrm{cite}(\cdot)\) lies outside the jurisdiction or effective interval of the task, and a majority-vote over top-\(n\) candidates stabilises the legal basis. The resulting citation pack \(\{\mathrm{cite}(l),\text{span}(l)\}_{l\in L(\tau)}\) is signed by the citation clerk and minted to the DAG.

Mapping legal text to matrix intent and exposure constitutes orientation rather than determination. A curator-assisted classifier proposes \(m(l)\in\{\text{Prohibition},\text{Obligation}\}\times\{\text{Internal},\text{External}\}\) for each chunk, with all proposals reviewed and versioned. At run time the router applies \(m(l)\) together with context to select \(P^{\star},O^{\star}\) and the order \(\prec\) (cf. \eqref{eq:feasible}, \eqref{eq:lexi}), while concrete guard templates and artefact schemas are compiled from the policy store. The complete retrieval and routing trace \(\langle v,e,\psi(\tau),L(\tau),\mathrm{cite}(\cdot)\rangle\) is minted to the DAG alongside the gate result, enabling replay and auditor verification.

This design addresses practical concerns. Legal aspects are not collapsed into a single global cluster but organised into multiple coordinated views. Agents are not tied one-to-one with statutes but are registered by capability and routed according to orientation and retrieved evidence. Correlation between chunks and laws is secured by canonical citations and effective dates rather than embedding heuristics. The outcome is portable across jurisdictions, resilient to textual drift, and ready for supervisory scrutiny.

\subsection{Evidence admissibility model}

Admissibility must be explicit to ensure that obligation checks in \eqref{eq:feasible}–\eqref{eq:lexi} are verifiable at run time and reproducible on audit. Each obligation \(o\in O^{\star}\) is modelled as producing an artefact bundle
\begin{align}
\mathrm{artefact}_o(\tau)
  &= \langle \text{text}, \text{hash}, \text{sig}, \text{prov}, \text{time}, \text{dag}, \text{ret} \rangle .
\end{align}

The admissibility predicate is defined as the conjunction of policy-specified quality dimensions,
\begin{equation}
E\bigl(\mathrm{artefact}_o(\tau)\bigr)\ \Leftrightarrow\ Q_{\mathrm{prov}}\ \wedge\ Q_{\mathrm{sig}}\ \wedge\ Q_{\mathrm{time}}\ \wedge\ Q_{\mathrm{dag}}\ \wedge\ Q_{\mathrm{ret}}\ \wedge\ Q_{\mathrm{access}},
\label{eq:evidence}
\end{equation}
where \(Q_{\mathrm{prov}}\) establishes provenance and chain of custody, \(Q_{\mathrm{sig}}\) validates authorised signatures and key material, \(Q_{\mathrm{time}}\) confirms trusted time binding, \(Q_{\mathrm{dag}}\) attests DAG inclusion through Merkle proof and index, \(Q_{\mathrm{ret}}\) enforces retention horizon and lawful basis, and \(Q_{\mathrm{access}}\) guarantees role-scoped readability with appropriate redaction of personal data.

Let \(\mathrm{Hash}(\cdot)\) denote a collision-resistant digest and \(\mathrm{TS}(\cdot)\) a trusted timestamp. A minimal admissible bundle must then satisfy
\begin{equation}
\text{hash}=\mathrm{Hash}(\text{text})\ \wedge\ \text{sig}=\mathrm{Sign}_{k_{\mathrm{oversight}}}(\text{hash})\ \wedge\ \text{time}=\mathrm{TS}(\text{hash})\ \wedge\ \text{dag}=\mathrm{Include}(\text{hash},\text{time}),
\label{eq:min-bundle}
\end{equation}
with \(\text{prov}\) linking to registered source identifiers and \(\text{ret}\) demonstrating that the object is scheduled for retention in accordance with policy and applicable law. Privacy is preserved by anchoring digests and metadata on the DAG while raw artefacts remain in controlled storage, accessible only through capability- and role-scoped URIs. This definition operationalises the obligation term in the feasible set \eqref{eq:feasible} and ensures that evidence produced by obligation checks is admissible by construction and audit-ready for supervisory verification.\\

\subsection{Policy compilation and governance}

The orientation matrix provides the initial signal, while policy compilation determines the active checks. Given the label \(v=(t,e)\) from \eqref{eq:rim-label} and the task context \(\chi(\tau)\) from \eqref{eq:context-check}, a compiler maps orientation and context into selected guards and their evaluation order, subject to human oversight and revision. The policy store is a versioned, time-scoped rule base with effective intervals, reviewer sign-off, and rollback support.

Compilation is modelled as a deterministic function of the effective policy version \(\nu\) at decision time \(t\):
\begin{equation}
\mathrm{Compile}_{\nu,t}\bigl(v,\text{context},\tau\bigr)\ \to\ \bigl(P^{\star},\,O^{\star},\,\prec,\,P^{\perp},\,B\bigr),
\label{eq:compile}
\end{equation}
where \(P^{\star}\subseteq P\) and \(O^{\star}\subseteq O\) are the selected prohibitions and obligations, \(\prec\) is a strict partial order over \(P^{\star}\cup O^{\star}\), \(P^{\perp}\subseteq P^{\star}\) are dominant disqualifiers, and \(B=\langle B_{\mathrm{risk}},B_{\mathrm{lat}},B_{\mathrm{cost}}\rangle\) are the active runtime budgets. Selection is constrained by jurisdiction and effective dates from the legal index, by the exposure label \(e\), and by any institutional policies specific to product or client tier. The matrix narrows the candidate set but does not enforce activation.

Budgets operate both at compile time and at run time. A budget-breach predicate informs committee activation in \eqref{eq:committee} and the externalisation gate in \eqref{eq:externalise}:
\begin{equation}
\mathrm{Breach}(\tau)\ :=\ \bigl(\mathrm{risk}(\tau) > B_{\mathrm{risk}}\bigr)\ \lor\ \bigl(\mathrm{latency}(\tau) > B_{\mathrm{lat}}\bigr)\ \lor\ \bigl(\mathrm{cost}(\tau) > B_{\mathrm{cost}}\bigr).
\label{eq:breach}
\end{equation}
Here risk may be a calibrated model score, while latency and cost are measured against defined service tiers. If the predicate fires, the compiler may escalate to a committee under \eqref{eq:committee} or deny externalisation under \eqref{eq:externalise}.

Obligation templates are bound to admissibility during compilation. For each \(o\in O^{\star}\), the compiler selects an evidence schema and admissibility test \(E(\cdot)\) from \eqref{eq:evidence}, producing a concrete template for \(\mathrm{artefact}_o(\tau)\) with required provenance, signatures, trusted time, and DAG inclusion. The feasible set \eqref{eq:feasible} and optimisation \eqref{eq:lexi} then operate over these compiled guards rather than the full candidate space.

Governance ensures reproducibility and institutional control. Each compilation yields a certificate
\begin{equation}
\mathrm{Cert}_{\mathrm{comp}}=\bigl\langle v,\,\mathrm{hash}(\text{context}),\,P^{\star},\,O^{\star},\,\prec,\,P^{\perp},\,B,\,\nu,\,\sigma_{\mathrm{compiler}},\,\sigma_{\mathrm{oversight}}\bigr\rangle,
\label{eq:cert}
\end{equation}
which is signed both by the compiler and by an oversight key, and anchored on the DAG with a reference to the policy store version. Replays reconstruct \(\mathrm{Cert}_{\mathrm{comp}}\) and verify that identical inputs under the same \(\nu\) yield identical selections and order.

Overrides are permitted under controlled conditions. An authorised reviewer may issue \(\mathrm{Override}(\mathrm{Cert}_{\mathrm{comp}},\Delta)\) that modifies \((P^{\star},O^{\star},\prec)\) with reason code and dual control. Safety monotonicity applies to dominant prohibitions: if \(p\in P^{\perp}\) and \(p(\pi,\tau)=1\), the override cannot deactivate \(p\) when exposure is external \((e=\text{External})\); any attempted weakening is logged and forces \(\textsc{deny}(c)\) under \eqref{eq:externalise}. Strengthening is always admissible, while removal of non-dominant checks requires reviewer signatures and the issuance of a new certificate version \(\nu'\).

Determinism and time anchoring close the loop. The compiler is deterministic given \((v,\text{context},\nu,t)\), and any change in output is attributable to context updates, policy version changes, or time-scoped rule amendments. Trusted time binds both \(\nu\) and \(\mathrm{Cert}_{\mathrm{comp}}\) to replayable evidence via \eqref{eq:min-bundle}. The result is a governance process that integrates admissibility, feasibility, committee formation, and externalisation, while keeping the orientation matrix as a guidance device rather than a source of hard-coded execution rules.

\subsection{Stylised simulation pseudocode}

The following pseudocode illustrates a pass through context detection, policy compilation, dominant disqualifier short-circuiting, admissibility checks, and the externalisation gate defined in Equation~\eqref{eq:externalise}.

\begin{algorithm}[H]
\caption{Execution flow for task $\tau$ (Part I)\protect\footnotemark}
\label{alg:task-execution-flow-part1}
\begin{algorithmic}[1]
\State \textbf{Input:} task $\tau$, context, candidate policy set $\Pi$, policy version $\nu$, time $t$
\If{$\chi(\tau) = 0$}
    \State $v \gets (\text{None}, \text{Internal})$
    \State MINT\_DAG($\tau$, decision = ``allow\_internal'')
    \State \Return allow\_internal
\Else
    \State $v \gets \text{ORIENT}(\text{context})$
    \State $(P^\star, O^\star, \prec, P^\bot, B) \gets \text{Compile}_{\nu,t}(v, \text{context}, \tau)$
    \If{$\text{Breach}(\tau, B)$}
        \State MINT\_DAG($\tau$, decision = ``deny'', reason = ``budget breach'')
        \State \Return deny(reason = ``budget breach'')
    \EndIf
    \ForAll{$p \in P^\bot$}
        \If{$\exists\, \pi \in \Pi \text{ such that } p(\pi, \tau) = 1$}
            \State MINT\_DAG($\tau$, decision = ``deny'', reason = ``dominant prohibition'')
            \State \Return deny(reason = ``dominant prohibition'')
        \EndIf
    \EndFor
\EndIf
\Statex \hfill\textit{(continued on next page)}
\end{algorithmic}
\end{algorithm}

\pagebreak

\begin{algorithm}[H]
\caption{Execution flow for task $\tau$ (Part II, continued)}
\label{alg:task-execution-flow-part2}
\begin{algorithmic}[1]
\setcounter{ALG@line}{19}   %
\If{$\chi(\tau) \neq 0$}
    \ForAll{$o \in O^\star$ in order $\prec$}
        \State $a \gets \text{EXECUTE}(o, \tau, \text{context})$
        \If{$E(a) = \text{false}$}
            \State MINT\_DAG($\tau$, decision = ``deny'', reason = ``inadmissible obligation'', failed = $o$)
            \State \Return deny(reason = ``inadmissible obligation'')
        \EndIf
    \EndFor
    \State $\Pi_{\text{feas}} \gets \{\pi \in \Pi \mid \forall p \in P^\star: p(\pi, \tau) = 0,\ \forall o \in O^\star: E(a) = \text{true}\}$
    \If{$\Pi_{\text{feas}} = \emptyset$}
        \State MINT\_DAG($\tau$, decision = ``deny'', reason = ``no feasible policy'')
        \State \Return deny(reason = ``no feasible policy'')
    \EndIf
    \State $\pi^\star \gets \arg\max_{\pi \in \Pi_{\text{feas}}} U(\pi, \tau)$
    \State MINT\_DAG($\tau$, decision = ``allow'', output = $\pi^\star$)
    \State \Return allow(output from $\pi^\star$)
\EndIf
\end{algorithmic}
\end{algorithm}

\footnotetext{Initial pseudocode generated using Claude Code in response to structured system prompts \cite{claudecode2024}.}

This pseudocode shows how the system detects applicability, assigns the orientation label, compiles policy deterministically, short-circuits on dominant disqualifiers, evaluates admissibility for obligation artefacts, optimises over the feasible set defined by \eqref{eq:feasible}–\eqref{eq:lexi}, executes the externalisation gate \eqref{eq:externalise}, and records every decision path by minting a replayable, time-bound audit pack.

\section{Discussion}

The architecture treats regulatory constraints as first-class objects while preserving the distinction between orientation and determination. The matrix provides a compact label \(v=(t,e)\) for classification and routing, while policy compilation selects and orders prohibitions and obligations subject to possible revision through oversight. Compliance is operationalised through feasibility and admissibility as defined in \eqref{eq:feasible}, \eqref{eq:lexi}, and \eqref{eq:externalise}, with every obligation bound to trusted time and anchored on a permissioned DAG to ensure replayability, provenance validation, and clear attribution of failure. A central benefit arises from the separation of concerns. Orientation remains stable and portable across jurisdictions, whereas activation is institution specific through the policy store and its governance processes. Prohibitions do not trade against utility, and obligations extend the task with artefacts that must satisfy the admissibility predicate \(E(\cdot)\) defined in \eqref{eq:evidence}. Committee activation under \eqref{eq:committee} ensures proportionality, while short-circuiting of dominant checks formalised in \eqref{eq:dominance} reduces latency without compromising the completeness of audit trails. There are trade-offs. Misclassification or an excessively broad dominance set may lead to false blocks, while incomplete policy stores can result in false releases. The legal index and the intent–exposure mapping \(m(l)\) therefore require continuous curation with versioning and reviewer sign-off. Privacy constraints further limit what can be anchored on the ledger; hashes and metadata are sufficient for attestation, but disciplined key management and controlled storage are required for raw artefacts. Determinism also relies on stable time sources and reproducible compilation given \((v,\text{context},\nu,t)\). The scope of the design is clearly delimited. It does not replace legal interpretation or product governance but instead provides an executable substrate that is law- and regulation-agnostic at the orientation layer and institution specific at the activation layer. Portability across DACH and the wider EU follows directly from this separation, since only compilation rules and evidence templates vary with local implementation.\\

\section{Conclusion}

The paper advances a compliance-first architecture that positions regulation as an orientation layer rather than a deterministic ruleset. The Regulatory Intent and Exposure Matrix provides the compact handle \(v=(t,e)\) for classification and routing, while execution is delegated to a governed policy compiler. Prohibitions constrain feasibility through the admissible set \(\Pi_{\mathrm{feas}}(\tau)\) in \eqref{eq:feasible} and block externalisation via \eqref{eq:externalise}. Obligations extend tasks with admissible artefacts that must satisfy the predicate \(E(\cdot)\) in \eqref{eq:evidence}. Optimisation then proceeds over \(\Pi_{\mathrm{feas}}(\tau)\) under lexicographic semantics in \eqref{eq:lexi}. Proportionality follows from policy-driven committee activation in \eqref{eq:committee}, while dominance rules in \eqref{eq:dominance} enable early termination without compromising audit completeness. Evidence, decisions and reason codes are bound to a permissioned DAG with deterministic timestamping, ensuring replayability, provenance checks and precise failure attribution.  Portability across DACH and the wider EU results from the separation between orientation and activation, a clause-level indexed legal corpus, and capability-based agent routing. The architecture transforms compliance from retrospective attestation into assurance by construction, offering a generalisable design pattern that remains law- and regulation-agnostic at the orientation layer while retaining institution-specific control at activation.\\

\section{Limitations and Future Research}
Several limitations remain. The design depends on curated mappings from legal text to orientation \(m(l)\to v\) and on the integrity of policy compilation; drift or gaps can cause false blocks or unintended releases. Formal guarantees for compiler determinism and reproducibility under \((v,\text{context},\nu,t)\) require explicit proof obligations and differential testing. The admissibility predicate \(E(\cdot)\) must be calibrated to institutional standards for provenance, signatures, trusted time and retention, and cryptographic choices and key management directly affect audit reliability. Ledger privacy remains constrained by linkage risks across hashes and metadata, and selective disclosure, salting and zero-knowledge proofs warrant further evaluation. Institutional dependencies also arise. Externalisation boundaries can shift when internal artefacts are reused in external contexts, which necessitates systematic exposure re-checks and reuse monitors. Committee thresholds \(k\) and runtime budgets \(B\) determine latency and cost, so empirical calibration, red-teaming and scenario replay are needed to quantify trade-offs and client impact. Retrieval robustness relies on similarity functions, feature maps and citation consistency, suggesting the need for adversarial tests and benchmarks specific to legal retrieval. Human and governance factors remain material. Curation of the legal index and the intent–exposure mapping requires ongoing reviewer effort with versioning and sign-off. Reviewer workload, override workflows and reason-code clarity all influence governance quality and practical adoption. Future research will extend evaluation on institution-specific scenarios in DACH, develop auditor-facing query suites tied to \eqref{eq:context-check}, \eqref{eq:rim-label}, \eqref{eq:feasible}, \eqref{eq:evidence} and \eqref{eq:externalise}, and investigate privacy-preserving ledger designs and time-binding mechanisms that improve verifiability without undermining data minimisation.

It should also be noted that empirical evaluation at meaningful scale would require full system deployment across institutional and regulatory contexts, which entails investments beyond the scope of academic research. Smaller prototypes would not capture the systemic properties of budget enforcement, admissibility, or externalisation, and could therefore provide misleading evidence. For this reason, the present contribution is deliberately theoretical and algorithmic: it establishes the formal and operational basis on which future large-scale empirical work can be responsibly pursued.

\bigskip

\bigskip

\printbibliography

@article{Kurz2025FMAAustria,
  author   = {Kurz, Walter and Magg, Reinhard and Stromeyer, Konrad},
  title    = {Financial and Operational Impacts of Regulatory Compliance on the Austrian Securities Industry},
  journal  = {Journal of Next-Generation Research 5.0},
  year     = {2025},
  volume   = {1},
  number   = {4},
  pages    = {137},
  doi      = {10.70792/jngr5.0.v1i5.137},
  issn     = {3075-2868},
  url      = {http://dx.doi.org/10.70792/jngr5.0.v1i5.137},
  language = {English}
}

@report{ecbFISEA2024,
  type         = {Report},
  author       = {{European Central Bank}},
  title        = {Financial integration and structure in the euro area},
  year         = {2024},
  institution  = {ECB},
  url          = {https://www.ecb.europa.eu/pub/pdf/fie/ecb.fie202406.en.pdf}
}

@report{bisIBS2025,
  type         = {Report},
  author       = {{Bank for International Settlements}},
  title        = {BIS international banking statistics and global liquidity indicators at end-September 2024},
  year         = {2025},
  institution  = {BIS},
  url          = {https://www.bis.org/statistics/rppb2501.pdf}
}

@misc{sifEU2025,
  author       = {{Swiss State Secretariat for International Finance (SIF)}},
  title        = {European Union: Market access and equivalence},
  year         = {2025},
  url          = {https://www.sif.admin.ch/en/european-union-eu}
}

@misc{crr2013,
  title        = {Regulation (EU) No 575/2013 on prudential requirements for credit institutions and investment firms ({CRR})},
  year         = {2013},
  organization = {Official Journal of the European Union},
  url          = {https://eur-lex.europa.eu/eli/reg/2013/575/oj}
}

@misc{crd2013,
  title        = {Directive 2013/36/EU on access to the activity of credit institutions and the prudential supervision of credit institutions ({CRD})},
  year         = {2013},
  organization = {Official Journal of the European Union},
  url          = {https://eur-lex.europa.eu/eli/dir/2013/36/oj}
}

@misc{mifid2_2014,
  title        = {Directive 2014/65/EU on markets in financial instruments ({MiFID~II})},
  year         = {2014},
  organization = {Official Journal of the European Union},
  url          = {https://eur-lex.europa.eu/eli/dir/2014/65/oj}
}

@misc{mifir_2014,
  title        = {Regulation (EU) No 600/2014 on markets in financial instruments ({MiFIR})},
  year         = {2014},
  organization = {Official Journal of the European Union},
  url          = {https://eur-lex.europa.eu/eli/reg/2014/600/oj}
}

@misc{solvency2_2009,
  title        = {Directive 2009/138/EC on the taking-up and pursuit of the business of insurance and reinsurance ({Solvency~II})},
  year         = {2009},
  organization = {Official Journal of the European Union},
  url          = {https://eur-lex.europa.eu/eli/dir/2009/138/oj}
}

@misc{ucits_2009,
  title        = {Directive 2009/65/EC on undertakings for collective investment in transferable securities ({UCITS})},
  year         = {2009},
  organization = {Official Journal of the European Union},
  url          = {https://eur-lex.europa.eu/eli/dir/2009/65/oj}
}

@misc{aifmd_2011,
  title        = {Directive 2011/61/EU on Alternative Investment Fund Managers ({AIFMD})},
  year         = {2011},
  organization = {Official Journal of the European Union},
  url          = {https://eur-lex.europa.eu/eli/dir/2011/61/oj}
}

@misc{emir_2012,
  title        = {Regulation (EU) No 648/2012 on OTC derivatives, central counterparties and trade repositories ({EMIR})},
  year         = {2012},
  organization = {Official Journal of the European Union},
  url          = {https://eur-lex.europa.eu/eli/reg/2012/648/oj}
}

@misc{csdr_2014,
  title        = {Regulation (EU) No 909/2014 on improving securities settlement in the European Union and on central securities depositories ({CSDR})},
  year         = {2014},
  organization = {Official Journal of the European Union},
  url          = {https://eur-lex.europa.eu/eli/reg/2014/909/oj}
}

@misc{psd2_2015,
  title        = {Directive (EU) 2015/2366 on payment services in the internal market ({PSD2})},
  year         = {2015},
  organization = {Official Journal of the European Union},
  url          = {https://eur-lex.europa.eu/eli/dir/2015/2366/oj}
}

@misc{emd2_2009,
  title        = {Directive 2009/110/EC on the taking up, pursuit and prudential supervision of the business of electronic money institutions ({EMD~II})},
  year         = {2009},
  organization = {Official Journal of the European Union},
  url          = {https://eur-lex.europa.eu/eli/dir/2009/110/oj}
}

@misc{amld4_2015,
  title        = {Directive (EU) 2015/849 on the prevention of the use of the financial system for the purposes of money laundering or terrorist financing (4{AMLD})},
  year         = {2015},
  organization = {Official Journal of the European Union},
  url          = {https://eur-lex.europa.eu/eli/dir/2015/849/oj}
}

@misc{amld5_2018,
  title        = {Directive (EU) 2018/843 amending Directive (EU) 2015/849 (5{AMLD})},
  year         = {2018},
  organization = {Official Journal of the European Union},
  url          = {https://eur-lex.europa.eu/eli/dir/2018/843/oj}
}

@misc{mica_2023,
  title        = {Regulation (EU) 2023/1114 on markets in crypto-assets ({MiCA})},
  year         = {2023},
  organization = {Official Journal of the European Union},
  url          = {https://eur-lex.europa.eu/eli/reg/2023/1114/oj}
}

@misc{dora_2022,
  title        = {Regulation (EU) 2022/2554 on digital operational resilience for the financial sector ({DORA})},
  year         = {2022},
  organization = {Official Journal of the European Union},
  url          = {https://eur-lex.europa.eu/eli/reg/2022/2554/oj}
}

@misc{gdpr_2016,
  title        = {Regulation (EU) 2016/679 on the protection of natural persons with regard to the processing of personal data ({GDPR})},
  year         = {2016},
  organization = {Official Journal of the European Union},
  url          = {https://eur-lex.europa.eu/eli/reg/2016/679/oj}
}

@misc{esmaArt16,
  author       = {{European Securities and Markets Authority}},
  title        = {MiFID~II, Article 16: Organisational requirements (Interactive Single Rulebook)},
  year         = {2023},
  url          = {https://www.esma.europa.eu/publications-and-data/interactive-single-rulebook/mifid-ii/article-16-organisational-requirements}
}

@misc{delReg2017_565,
  title        = {Commission Delegated Regulation (EU) 2017/565 supplementing Directive 2014/65/EU},
  year         = {2017},
  organization = {Official Journal of the European Union},
  url          = {https://eur-lex.europa.eu/legal-content/EN/TXT/PDF/?uri=CELEX:32017R0565}
}

@misc{its2021451,
  title        = {Commission Implementing Regulation (EU) 2021/451 laying down {ITS} for supervisory reporting under the {CRR}},
  year         = {2021},
  organization = {Official Journal of the European Union},
  url          = {https://eur-lex.europa.eu/eli/reg_impl/2021/451/oj}
}

@misc{esmaArt34,
  author       = {{European Securities and Markets Authority}},
  title        = {MiFID~II, Article 34: Freedom to provide investment services and activities (passporting)},
  year         = {2023},
  url          = {https://www.esma.europa.eu/publications-and-data/interactive-single-rulebook/mifid-ii/article-34-freedom-provide-investment}
}

@misc{ebaPSD2Art28,
  author       = {{European Banking Authority}},
  title        = {PSD2, Article 28: Application to exercise the right of establishment and freedom to provide services},
  year         = {2024},
  url          = {https://www.eba.europa.eu/regulation-and-policy/single-rulebook/interactive-single-rulebook/16236}
}

@misc{germany_kwg,
  title        = {Gesetz über das Kreditwesen ({KWG})},
  year         = {2025},
  organization = {Gesetze im Internet (BMJ/Bundesanzeiger Verlag)},
  url          = {https://www.gesetze-im-internet.de/kredwg/}
}

@misc{germany_wphg,
  title        = {Wertpapierhandelsgesetz ({WpHG})},
  year         = {2025},
  organization = {Gesetze im Internet (BMJ/Bundesanzeiger Verlag)},
  url          = {https://www.gesetze-im-internet.de/wphg/}
}

@misc{bafin_wphg,
  author       = {{BaFin}},
  title        = {Securities Trading Act ({WpHG})—overview},
  year         = {2019},
  url          = {https://www.bafin.de/SharedDocs/Veroeffentlichungen/EN/Aufsichtsrecht/Gesetz/WpHG_en.html}
}

@misc{austria_bwg,
  title        = {Bankwesengesetz ({BWG})},
  year         = {2025},
  organization = {RIS Bundeskanzleramt},
  url          = {https://www.ris.bka.gv.at/GeltendeFassung.wxe?Abfrage=Bundesnormen&Gesetzesnummer=10004827}
}

@misc{austria_wag2018,
  author       = {{Finanzmarktaufsicht (FMA)}},
  title        = {Securities Supervision Act 2018 ({WAG} 2018) [convenience translation]},
  year         = {2024},
  url          = {https://www.fma.gv.at/wp-content/plugins/dw-fma/download.php?d=2085},
  note         = {Check against official RIS version}
}

@misc{fmaIntegrated,
  author       = {{Finanzmarktaufsicht (FMA)}},
  title        = {Financial market supervision in Austria (integrated model)},
  year         = {2025},
  url          = {https://www.fma.gv.at/en/financial-market-supervision-in-austria/}
}

@misc{fedlex_finsa,
  title        = {Federal Act on Financial Services ({FinSA})},
  year         = {2024},
  organization = {Fedlex},
  url          = {https://www.fedlex.admin.ch/eli/cc/2019/758/en}
}

@misc{fedlex_finia,
  title        = {Federal Act on Financial Institutions ({FinIA})},
  year         = {2024},
  organization = {Fedlex},
  url          = {https://www.fedlex.admin.ch/eli/cc/2018/801/en}
}

@misc{finmaLegalBasis,
  author       = {{FINMA}},
  title        = {FINMA’s legal basis (Financial Market Supervision Act and sectoral laws)},
  year         = {2025},
  url          = {https://www.finma.ch/en/documentation/finma-s-legal-basis/}
}

@misc{ec_equivalence,
  author       = {{European Commission}},
  title        = {Equivalence of non-EU financial frameworks: overview and list of decisions},
  year         = {2025},
  url          = {https://finance.ec.europa.eu/eu-and-world/equivalence-non-eu-financial-frameworks_en}
}

@report{efamaFactbook2025,
  type         = {Report},
  author       = {{EFAMA}},
  title        = {Trends in European investment funds: Fact Book 2025},
  year         = {2025},
  institution  = {European Fund and Asset Management Association},
  url          = {https://www.efama.org/sites/default/files/fact-book-2025_lowres.pdf},
}

@report{eibis2024Austria,
  type         = {Report},
  author       = {{European Investment Bank}},
  title        = {EIB Investment Survey 2024: Country overview—Austria},
  year         = {2025},
  institution  = {EIB},
  url          = {https://www.eib.org/files/documents/lucalli/20240238_econ_eibis_2024_austria_en.pdf}
}

@report{bisfsi2018Proportionality,
  type         = {Report},
  author       = {{Financial Stability Institute}},
  title        = {Proportionality in banking regulation: a cross-country comparison},
  year         = {2018},
  institution  = {Bank for International Settlements},
  url          = {https://www.bis.org/fsi/publ/insights1.pdf}
}

@report{fsb2019SME,
  type         = {Report},
  author       = {{Financial Stability Board}},
  title        = {Evaluation of the effects of financial regulatory reforms on {SME} financing},
  year         = {2019},
  institution  = {FSB},
  url          = {https://www.fsb.org/wp-content/uploads/P291119-1.pdf}
}

@article{Sun2025MultiAgentHDRL,
  author  = {Sun, R.},
  title   = {A novel multi-agent hierarchical deep reinforcement learning framework for portfolio optimization},
  journal = {Journal of Finance and Computation},
  year    = {2025},
}

@article{Cheng2024MultiagentRL,
  author  = {Cheng, L. C.},
  title   = {Multi-agent based deep reinforcement learning framework for trading strategy and portfolio management},
  journal = {Applied Artificial Intelligence in Finance},
  year    = {2024},
}

@article{Hambly2021RLFinance,
  author  = {Hambly, B. and Xu, R. and Yang, H.},
  title   = {Recent advances in reinforcement learning in finance},
  journal = {Mathematical Finance},
  year    = {2021},
  volume  = {31},
  number  = {3},
  pages   = {782--813},
  doi     = {10.1111/mafi.12315}
}

@article{Malibari2023RLFintech,
  author  = {Malibari, N. and Katib, I. and Mehmood, R.},
  title   = {Systematic review on reinforcement learning in the field of FinTech},
  journal = {Journal of Cloud Computing},
  year    = {2023},
  volume  = {12},
  number  = {1},
  pages   = {1--34},
  doi     = {10.1186/s13677-023-00449-1}
}

@article{Ali2021AuditLogSurvey,
  author  = {Ali, A. and Ahmed, M. and Khan, A.},
  title   = {Audit logs management and security: a survey},
  journal = {Security and Privacy},
  year    = {2021},
  volume  = {4},
  number  = {3},
  pages   = {e155},
  doi     = {10.1002/spy2.155}
}

@article{IJITMIS2025BPMAudit,
  author  = {Singh, P. and Kumar, A.},
  title   = {Ensuring data integrity and auditability in BPM systems: A blockchain-based approach},
  journal = {International Journal of Information Technology \& Management Information Systems},
  year    = {2025},
}

@article{Faccia2022PermissionedAudit,
  author  = {Faccia, A.},
  title   = {Is permissioned blockchain the key to support the external audit process?},
  journal = {International Journal of Disclosure and Governance},
  year    = {2022},
  volume  = {19},
  number  = {4},
  pages   = {354--364},
  doi     = {10.1057/s41310-022-00153-2}
}

@article{Arham2025AIBlockchainAudit,
  author  = {Arham, M. W.},
  title   = {Transforming auditing through AI and blockchain},
  journal = {American Journal of Industrial and Business Management},
  year    = {2025},
  volume  = {15},
  number  = {2},
  pages   = {225--241},
  doi     = {10.4236/ajibm.2025.152011}
}

@misc{DORA_EIOPA2025,
  author       = {{EIOPA}},
  title        = {Digital Operational Resilience Act ({DORA})},
  year         = {2025},
  howpublished = {\url{https://www.eiopa.europa.eu/digital-operational-resilience-act-dora_en}},
  note         = {Overview of DORA measures and scope}
}

@misc{DORA_Skadden2024,
  author       = {{Skadden}},
  title        = {The EU's Digital Operational Resilience Act ({DORA})},
  year         = {2024},
  howpublished = {\url{https://www.skadden.com/insights/publications/2024/07/the-eus-digital-operational-resilience-act}},
  note         = {Analysis of ICT risk requirements under DORA}
}

@misc{EBA_RTS2024,
  author       = {{European Banking Authority}},
  title        = {Regulatory Technical Standards on ICT services supporting critical or important functions},
  year         = {2024},
  howpublished = {\url{https://www.eba.europa.eu/activities/single-rulebook/regulatory-activities/operational-resilience/regulatory-technical-standards-policy-ict-services-supporting-critical-or-important-functions}},
  note         = {DORA RTS on ICT third-party risk governance}
}

@misc{claudecode2024,
  author       = {{Anthropic}},
  title        = {Claude Code},
  year         = {2024},
  howpublished = {\url{https://www.anthropic.com/index/claude-code}},
  note         = {Accessed 17 August 2025}
}

\swDeclarationsPage

\end{document}